\documentclass[aps,prl,reprint,amsmath,amssymb,
superscriptaddress,showpacs,showkeys]{revtex4-1}
\usepackage{graphicx}
\usepackage{dcolumn}
\usepackage{bm}
\usepackage{soul} 
\usepackage{color}
\definecolor{vs}{rgb}{0.1,0.4,0.1}                  

\usepackage{hyperref}
\usepackage{verbatim}
\newcommand\wordcount{\verbatiminput{\jobname.sum}}
\begin{document}

\title{Dynamic Fano resonances: From toy model to resonant Mie scattering}

\author{Michael I. Tribelsky}
\email[Corresponding author (replace ``\_at\_" by @):\\]{\mbox{E-mail: mitribelsky\_at\_gmail.com}
}
\homepage[]{https://polly.phys.msu.ru/en/labs/Tribelsky/}
\affiliation{
M. V. Lomonosov Moscow State University, Moscow, 119991, Russia}
\affiliation{National Research Nuclear University MEPhI (Moscow Engineering Physics Institute), Moscow, 115409, Russia}
\affiliation{Landau Institute for Theoretical Physics RAS, Chernogolovka, Moscow Region 142432, Russia}
\affiliation{RITS Yamaguchi University, Yamaguchi, 753-8511, Japan}

\author{Andrey E. Miroshnichenko}
\email[E-mail: ]
{\mbox{andrey.miroshnichenko\_at\_unsw.edu.au}}
\affiliation{School of Engineering and Information Technology, University of New South Wales, Canberra, ACT, 2600, Australia}

\date{\today}

\begin{abstract}
Based on the substantial difference in the response time for the resonant and background partitions at stepwise variations of the exiting signal, a simple exactly integrable model describing the dynamic Fano resonance (DFRs) is proposed. The model does not have any fitting parameters, may include any number of resonant partitions and exhibits high accuracy. It is shown that at the point of the destructive interference any sharp variation of the amplitude of the excitation (no matter an increase or a decrease) gives rise to pronounced ``flashes" in the intensity of the output signal. In particular, the flash should appear behind the trailing edge of the exciting pulse, when the excitation is already over. The model is applied to explain the DFRs at the light scattering by a dielectric cylinder with two resonant modes excited simultaneously and exhibits the excellent agreement with the results of the direct numerical integration of the Maxwell equations.
\end{abstract}

\pacs{42.25.Fx, 42.65.Es, 46.40.Ff, 78.67.Bf}

\maketitle
%
%
\emph{Introduction.} It is impossible to overrate the importance of the Fano resonances --- the original paper of Ugo Fano~\cite{Fano:PR:1961} is one of the most cited papers ever published in all journals of the Physical Review series. Nowadays, the phenomenon is more important than ever owing to its numerous applications in nanophysics and nanotechnologies~\cite{Lukyanchuk2010,Miroshnichenko2010,Jacob:NN:2016}.

Recently, the frontier of modern physic has moved toward ultrafast processes with the time-scale comparable with the atomic relaxation times. As any resonant phenomenon, the Fano resonances have a certain response time to react to sharp changes in the exciting signal. It brings about new interesting time-dependent effects. These effects are extensively explored in atomic spectroscopy both theoretically and experimentally~\cite{ott2013lorentz,kaldun2016observing,gruson2016attosecond}.

In optics, different transient effects in open resonant structures are well-known too~\cite{sirenko2007modeling}. However, we are not aware of the study of the DFRs at the resonant Mie scattering by particles. Meanwhile, on the one hand, the time-scale of the corresponding dynamics lies within the resolution of the modern experimental technique. On the other hand, the results in quantum physics obtained for the wave function cannot be directly applied to the solutions of the Maxwell equations. Thus, the study of the DFRs at the resonant Mie scattering may bring about new effects which do not exist in the steady-state scattering and therefore is highly desirable.

Moreover, since the steady-state Fano resonances are observed in a wide diversity of physical systems, the same should be true for the DFRs. Then, a simple, yet accurate model capturing all generic features of the DRFs and free from specific peculiarities of a given problem, i.e., a model applicable to the description of the DFRs in \emph{any} case of their occurrence is of great demand.

In the present Letter we try to respond to the challenges, providing a new insight to the problem.
Specifically, first, we revisit the model of the forced oscillations of weakly coupled harmonic oscillators~\cite{Golovinski:ArXiv2017}. We show that the model may be reduced to the forced oscillations of a \emph{single} oscillator with a complex eigenfrequency. The result of this analysis is the identification of the generic features of the DFRs and separation them from the individual peculiarities of the given system. Then, we consider the DFRs in the actual physical problem of light scattering by an infinite circular dielectric cylinder. The obtained generic features of the phenomenon help us to select the values of the refractive index and the radius of the cylinder so that the manifestation of the DFRs is the most pronounced. The corresponding value of the refractive index equal to 3.125. It coincides with that for typical semiconductors, such as Si, GaAs, GaP, in NIR spectral domain~\cite{refractiveindex.info}.

The light scattering problem was inspected with the help of direct numerical integration of the Maxwell equations. Next, based on the analysis of the dynamics of the coupled oscillators a simple integrable model \emph{without any fitting parameter} is proposed to describe the DFRs observed in the simulation. The comparison of the results obtained within the framework of the model with the numerics exhibits the excellent accuracy of the model. Finally, in the conclusions, we summarize and highlight the main results of the study.

\emph{Coupled Oscillators vs Temporal Coupled-Mode Theory.} There are just two model approaches to the Fano resonances based on time-dependent differential equations and therefore, at least in principle, making possible to study the DFRs, namely the aforementioned model of the driven coupled oscillators~\cite{Joe2006,Golovinski:ArXiv2017,Tribelsky:RITS} and the  Temporal Coupled-Mode Theory (TCMT)~\cite{louisell1960coupled,Fan:PRL:2010}. While, for the time being, the TCMT is widely used for the quantitative description of the steady-state resonant light scattering~~\cite{Verslegers:10,Fan:jp9089722,Soljacic:nl500340n}, to the best of our knowledge, it has not been applied to inspect the actual transient processes in this field. In contrast, while the DFRs for the coupled classical oscillators have been discussed~\cite{Golovinski:ArXiv2017} the extension of these results to the resonant light scattering in a wide variety of actual physical systems is still missing.

Note, despite all its power, TCMT is not quite suitable to treat the DFRs. Among other disadvantages its application to this problem exhibits the following intrinsic inconsistency --- one of the coefficients employed in TCMT is the ratio of the instantaneous amplitudes of the scattered and incident pulses. Meanwhile, the most interesting, counterintuitive transient effect exhibiting by the DFRs is a ``flash" of the scattered radiation after the incident pulse is over, explained by the irradiation of electromagnetic energy accumulated in the scattering particle in the previous stages of the scattering process, see below. During the flash, the amplitude of the incident wave is zero, while the one for the scattered wave is still finite. It results in the divergence of the corresponding coefficient, which makes the TCMT inapplicable. For this reason in what follows we focus on the model of coupled oscillators.

\emph{The model}. We begin with the conventional dynamical model of two coupled harmonic oscillators with complex coordinates $z_{1,2}(t)$. The first oscillator is excited by the driving force $A(t)\exp[-i\omega t]$ and coupled with the second one. The first oscillator is dissipative with the damping constant $2\gamma$. The second oscillator does not have any intrinsic dissipation.

For the proceeding analysis, it is convenient to employ the linearity of the problem, presenting $z_1$ as a sum of two terms: $z_1 = z_{11} + z_{12}$ and to write the governing equations in the form
\begin{eqnarray}
  & & \ddot{z}_{11} + 2\gamma\dot{z}_{11} +\omega_{1}^2 z_{11} = A(t)\exp[-i\omega t], \label{eq:z11} \\
  & & \ddot{z}_{12} + 2\gamma\dot{z}_{12} +\omega_{1}^2 z_{12} = \kappa z_2,  \label{eq:z12}\\
  & & \ddot{z}_2 + \omega_{2}^2 z_2 = \kappa(z_{11}+z_{12}), \label{eq:z2}
  \end{eqnarray}
supplemented with the initial conditions $\dot{z}_{11}(0)={z}_{11}(0)=\dot{z}_{12}(0)={z}_{12}(0)=\dot{z}_{2}(0)={z}_{2}(0) = 0$. Here dot stands for $d/dt$.

Eq.~\eqref{eq:z11} is the well-known, exactly integrable equation for a single driven oscillator. Let us proceed with the analysis of the remaining two equations.


In contrast to the spike (the Dirac-delta-function) excitation discussed in the quantum manifestation of the DFRs~\cite{ott2013lorentz,kaldun2016observing} we consider a rectangular driving pulse with $A(t)$ equals $A_0=const$ at \mbox{$0 \leq t \leq \tau$} and zero outside this domain. At $\tau$ larger than the transient period this choice of $A(t)$ makes it possible studying the ``pure" complete transient process to the steady-state scattering at the leading edge of the exciting pulse and the decay of the steady-state scattering at its trailing edge.

It is known that at $A = const$ the $\omega$-dependence of the amplitude of $z_1$ at the steady-state oscillations exhibits the conventional Fano profile with the complete vanishing of $z_1$ at $\omega = \omega_{2}$~\cite{Joe2006,Tribelsky:RITS}, see also the Supplemental Material~I~\footnote{See the Supplemental Material I at http://link.aps.org/supplemental/???/PhysRevLett.??? for details of routine cumbersome calculations.}. In what follows we discuss the case of weak dissipation ($\gamma \ll \omega$) and weak coupling (the quantitative restriction for $\kappa$ will be formulated below).

The destructive interference at the Fano resonances happens owing to the superposition of the two partitions ---  resonant and background (nonresonant) ones at the point where they have equal amplitudes and opposite phases~\cite{Fano:PR:1961,Miroshnichenko2010,Tribelsky:RITS}. It gives rise to the vanishing of the output signal. In our model it is $z_1$. Then, what are the resonant and background partitions? The difference between the partitions is in their $\omega$-dependence --- for the background partition, this dependence is weak, while for the resonant partition it is sharp. Let us inspect the lineshapes for $|z_{11}|^2$ and $|z_{12}|^2$ at the steady-state oscillations in the vicinity of $\omega = \omega_{2}$. The corresponding analysis is trivial but cumbersome, see the Supplemental Material I~\cite{Note1}. Its results are as follows: The line for $|z_{11}|^2$ is the usual harmonic oscillator line the linewidth $\Gamma_1 = 2\gamma$.

Regarding $|z_{12}|^2$ and $|z_2|^2$, both have the conventional Lorentzian lineshape with different amplitudes but the same linewidth:
\begin{equation}\label{eq:gamma_eff}
  \Gamma_{\rm eff}=2\gamma_{\rm eff};\;\gamma_{\rm eff}=\frac{\gamma\kappa^2}{4(\omega_{0}^2\Delta\omega_{12}^2+\gamma^2\omega_{2}^2)},
\end{equation}
where $\omega_0 = (\omega_{1}+\omega_{2})/2,\; \Delta\omega_{12} = \omega_{2}-\omega_{1}$. The weak coupling corresponds to the case $\gamma_{\rm eff} \ll \gamma$. Then, according to the general Fano concept~\cite{Fano:PR:1961}, $z_{11}$ should be regarded as the background partition, while $z_{12}$ is the resonant one.

However, there is another important conclusion following from the inequality $\gamma_{\rm eff} \ll \gamma$. The characteristic time scale determining the \emph{intrinsic dynamical response} of a damped harmonic oscillator to any external perturbation is $1/\gamma$. This is true for $z_{11}$. Regarding the second oscillator, it does not have its own dissipation. Its relaxation is determined solely by the energy transfer to the first oscillator though their coupling. The corresponding relaxation time must diverge at $\kappa \rightarrow 0$. Then, it may be expected that at small enough $\kappa$ this time is much larger than $1/\gamma$.

As for $z_{12}$, Eq.~\eqref{eq:z12} is the conventional equation of forced oscillations for an oscillator with the damping constant $2\gamma$ and driving force $\kappa z_2$. Then, the characteristic response time for $z_{12}$ to a perturbation is the same as that for $z_{11}$, i.e., $1/\gamma$, which is much smaller than the one for $z_2$. It means $z_{12}$ follows slow variations of the amplitude of $z_2$ \emph{adiabatically}. The obtained narrow linewidth for $|z_{12}|^2$ has nothing to do with its relaxation time. This is just the profile of the corresponding effective driving force $|\kappa z_2|^2$, which $|z_{12}|^2$ reproduces owing to the linearity of the problem and the ``adiabatic connections" of $z_{12}$ with $z_2$.

In this case, the temporal dependence of $z_{12}$ is given by the well-known steady-state solution for a single driven oscillator, where the amplitude of the driving force should be replaced by $\kappa z_2(t)$. Putting this expression together with that for $z_{11}(t)$, obtained by the integration of Eq.~\eqref{eq:z11}, into Eq.~\eqref{eq:z2} brings about the \emph{decoupled} equation for $z_2$. This is the usual equation for the forced harmonic oscillations of a \emph{single} oscillator. However, despite the damping factor in the equation is zero, the oscillator has the complex eigenfrequency:
\begin{equation}\label{eq:complex_omega2}
\widetilde{\omega}_2 = \omega_2\left(1+\frac{\kappa^2}{\omega_2^2(\omega^2-\omega_1^2+2i\gamma\omega)}\right)^{1/2},
\end{equation}
resulting in the exponential decay of the amplitude of the free oscillations. The obtained equation for $z_2$ is also exactly integrable. Its solution is the undamped forced oscillations with the frequency $\omega$ superimposed with the decaying free oscillations with the frequency equal to Re$\,\widetilde{\omega}_2$ and the decrement Im$\,\widetilde{\omega}_2$. Expanding the square root in Eq.~\eqref{eq:complex_omega2} in powers of small ${\kappa^2}$ we obtain that in the most interesting case of the complete destructive interference, i.e., at $\omega = \omega_2$ the decrement for the free oscillations is exactly the same $\gamma_{\rm eff}$, see Eq.~\eqref{eq:gamma_eff}.

Then, neglecting the fast transient of the background partition to the steady-state oscillations, we can say that \mbox{$|z_{11}(t)+z_{12}(t)|^2$} exhibits the pure exponential decay with the decrement exactly equal to the linewidth of the resonant partition at the steady-state oscillations. The corresponding analysis is trivial. We put it into the Supplemental Material~I~\cite{Note1}. Note, that $z_{11}(t)$ achieves the steady state in the time $\sim 1/\gamma$ and in the same time $z_{12}(t)$ achieves the quasi-steady state (the adiabatic connection with $z_2(t)$). Since that moment the phase shift between $z_{11}(t)$ and $z_{12}(t)$ is fixed and the relaxation to the complete destructive interference occurs solely owing to the growth of the \emph{amplitude} of the resonant partition.
%

The dynamics of the decay of the oscillations at \mbox{$t \geq \tau$} may be treated in a similar manner. The treatment gives rise to the following scenario: The fast damping of the free oscillations of $z_{11}(t)$ (background) with the characteristic time scale $1/\gamma$ occurs while the amplitude of the resonant partition $z_{12}(t)$ remains practically unchanged. It destroys the balance between the resonant and background partitions required for the destructive interference. As a result, the amplitude of the output signal \mbox{$z_1 = z_{11}+z_{12}$} \emph{sharply increases} up to the amplitude of the steady-state oscillations of the resonant partition. Then, a slow pure exponential vanishing of \mbox{$|z_1|^2 \cong |z_{12}|^2$} with the decrement $\Gamma_{\rm eff}$ takes place. At $\gamma_{\rm eff}/\gamma \ll 1$ the accuracy of the obtained approximate solution is high and increases with a decrease in this ratio~\cite{Note1}.

\emph{Generic vs specific}. Let us try to understand, which of the results obtained are generic and which are specific just for the system discussed. By the definition, the $\omega$-dependence of the steady-state amplitude of the resonant partition must have a sharp pronounces maximum at a resonant frequency. In the vicinity of the maximum, such a profile always may be approximated by a Lorentzian and hence the dynamics of the corresponding resonant mode may be described by the dynamics of a \emph{single} driven oscillator with the characteristic response time equal to the inverse linewidth, no matter what the actual physical nature of this mode is. Obviously, such a property is generic.

In contrast, again from the definition, the $\omega$-dependence of the steady-state amplitude of a background partition is weak. It means, the dynamics of the response of the background partition to external perturbations is much faster than that for the resonant one. However, the details of this fast dynamics substantially depend on the specific features of a specific system in question.

Fortunately, these details are unimportant, if the slow dynamics is the only concern. In this case, the fast-dynamic-stage may be just neglected. Then, the assumption that at an abrupt variation of the amplitude of the external perturbation the background partitions achieve the new equilibrium states immediately, while the amplitudes of the resonant partitions remain unchanged is a very good approximation to the actual process. This state may be regarded as the initial conditions for the slow relaxation of the resonant partitions.

A generic and rather unexpected conclusion is that in any system exhibiting the Fano resonances any abrupt change in the amplitude of the external (incoming) signal, no matter an increase or a decrease, if it happens at the point of the destructive interference, must result in \emph{a sharp increase} in the amplitude of the output signal. The relaxation of the increase to the new equilibrium state obeys the exponential law with the decrement equal to the half-linewidth of the resonant partition.

\emph{DFRs in light scattering by a particle.} To illustrate the power of the developed approach we apply it to DFRs at the resonant Mie scattering. For the sake of simplicity we consider the light scattering by an infinite circular lossless dielectric cylinder with the base radius $R$ irradiated in a vacuum by a plane linearly polarized electromagnetic wave whose wave vector $\mathbf{k}$ is perpendicular to the cylinder axis and vector $\mathbf{E}$ oscillates in the plane of the base (TE polarization, normal incidence). To make this problem similar to the one discussed above we also consider the same rectangular pulse with duration $\tau$ and frequency $\omega$. It is supposed that $\tau$ is much larger any other time scales of the problem so that the wave inside the pulse may be regarded as monochromatic.

The steady-state version of the problem is exactly solvable, see, e.g.~\cite{bohren2008absorption}. According to this solution, the fields in the Maxwell equations are presented as the infinite series of partial waves (multipoles). The amplitudes of the partial waves are proportional to the so-called scattering coefficients. For the problem in question, there are just two sets of these coefficients: $a_\ell$ (describing the scattered field outside the cylinder) and $d_\ell$ (for the field within it). Here $\ell$ designates the multipolarity ($-\infty < \ell < \infty$). The scattering coefficients are expressed in terms of the Bessel functions. The corresponding formulae are well-known~\cite{bohren2008absorption}. We put them in the Supplemental Material I~\cite{Note1} for references.

To characterize the dynamics of the light scattering we introduce the instantaneous scattering efficiency per the unit of length of the cylinder axis $Q_{\rm sca}(t)$. The only difference between the conventional scattering efficiency~\cite{bohren2008absorption} and $Q_{\rm sca}(t)$ is that we do not perform the averaging of the Poynting vector over the period of the field oscillations --- the Poynting vector is defined as a real quantity proportional to the vector product of the real parts of the instantaneous values of fields $\mathbf{E(\mbox{$t$},r)}$ and $\mathbf{H(\mbox{$t$},r)}$ in a given point of the space $\mathbf{r}$.

For the steady-state scattering the routine calculations, employing the asymptotical behavior of the scattered field at large $r$, analogous to the conventional ones~\cite{bohren2008absorption} give rise to the following expression for $Q_{\rm sca}(t)$:
\begin{eqnarray}
   & & Q_{\rm sca}=\sum_{\ell=-\infty}^{\infty}Q_{{\rm sca}\,(\ell)};\; Q_{{\rm sca}\,(\ell)} = Q_{{\rm sca}\,(\ell)}^{(0)} + Q_{{\rm sca}\,(\ell)}^{(\rm osc)}, \label{eq:Qsca} \\
   & & Q_{{\rm sca}\,(\ell)}^{(0)}\! =\! \frac{2}{q}|a_\ell(t)|^2;\; Q_{{\rm sca}\,(\ell)}^{(\rm osc)}\! = \! -\frac{i}{q}\!\left[a_\ell^2(t) e^{2ikr}\!+\!c.c.\right]\!\!, \label{eq:Q_through_a}
\end{eqnarray}
where $q = kR$ is the size parameter, $k=\omega/c$ stands for the wavenumber of the incident wave, $c$ is the speed of light in a vacuum, $a_\ell(t) = a_\ell\exp[-i\omega t]$, and $a_\ell$ are the conventional scattering coefficients~\cite{bohren2008absorption,Note1}.

Coefficients $a_\ell$ and $d_\ell$ are connected with each other by the identity~\cite{Tribelsky_Mirosh_2016}:
\begin{equation}\label{eq:al-dl}
  a_\ell \equiv a_\ell^{(\rm PEC)} - \frac{J^\prime_\ell(mq)}{H^{(1)\prime}_\ell(q)}d_\ell;\; a_\ell^{(\rm PEC)} \equiv \frac{J^\prime_\ell(q)}{H^{(1)\prime}_\ell(q)},
\end{equation}
where $m$ is the refractive index of the cylinder, $J_\ell(z)$ and $H^{(1)}_\ell(z)$ stand for the Bessel and Hankel functions, respectively, prime denotes the derivative over the entire argument of a function, and $ a_\ell^{(\rm PEC)}$ is the scattering coefficient of the same cylinder made of the so-called $P$erfect $E$lectric $C$onductor.

As it is shown in Ref.~\cite{Tribelsky_Mirosh_2016} for the Fano resonances exhibiting by high-index particles $a_\ell^{(\rm PEC)}$ plays the role of the background partition ($z_{11}$ in our model), the second term in the identity Eq.~\eqref{eq:al-dl} is the resonant one ($z_{12}$), $d_\ell$ should be regarded as $z_2$, and  $-{J^\prime_\ell(mq)}/{H^{(1)\prime}_\ell(q)}$  is the constant in the expression $z_{12}(t) \cong const\cdot z_2(t)$~\cite{Note1}.

\begin{figure}
  \centering
  \includegraphics[width=.5\textwidth]{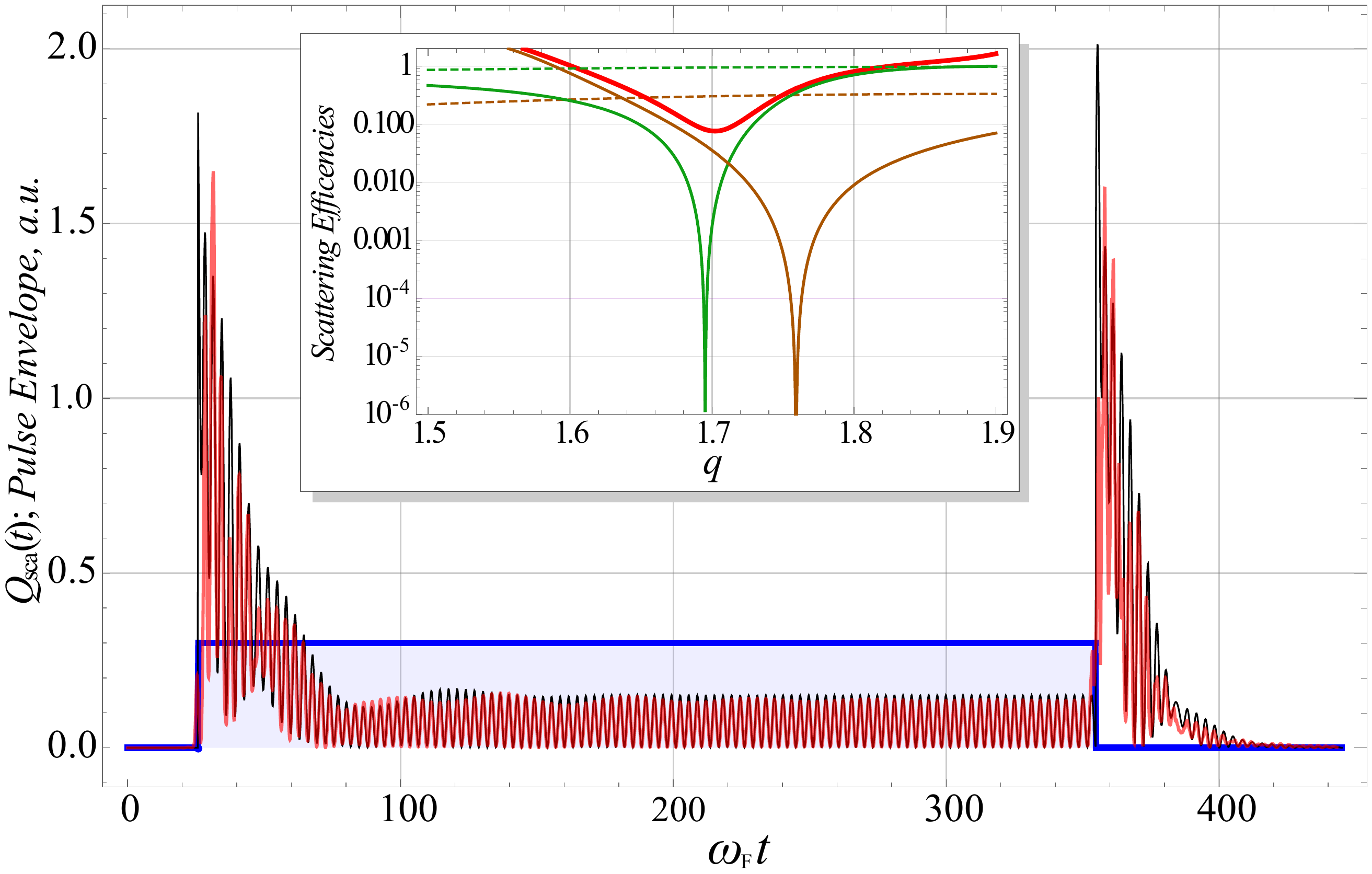}
  \caption{$Q_{\rm sca}(t)$ obtained by the direct numerical integration of the Maxwell equations (red) and its comparison with the model of the driven oscillators (black). The envelope of the incident pulse (in a.u.) is shown in blue. To compensate the time required for the light to cover the distance from the inlet to the cylinder and from the one to the monitor the incident pulse is shifted so that the leading edges of the incident and scattered pulses coincide. Note the sharp intensive flashes of the scattered radiation at the beginning of the scattering process and after the incident pulse is over. The inset shows the steady-state scattering efficiency $Q_{\rm sca}^{(0)}$ (thick, red), partial efficiencies $Q_{\rm sca\;(0)}^{(0)}$ (green, solid), $Q_{\rm sca\;(2)}^{(0)}+Q_{\rm sca\;(-2)}^{(0)}$ (brown, solid) and the same quantities for the corresponding PEC modes (dashed lines), as functions of the size parameter $q$.}\label{fig:Q0}
\end{figure}

To have the discussed effects the most pronounced we need to find such values of $q$ and $m$ that in the close vicinity of them (i) the corresponding partial waves exhibit the destructive Fano interference, (ii) the values $|a_\ell^{(\rm PEC)}|^2$ are as large as possible, and (iii) the contribution of the off-resonant multipoles to the steady-state scattering efficiency are minimal. For the problem in question good candidates are the pair \mbox{$m = m_{_F} \cong 3.125$}, \mbox{$q = q_{_F} \cong 1.702$.} The pair corresponds to the local minimum of $Q_{\rm sca}^{(0)} = \sum Q_{{\rm sca}\,(\ell)}^{(0)}=0.0760...$ The minimum is related to the fact that in the close vicinity of this point the two partial efficiencies, $Q_{{\rm sca}\,(0)}^{(0)}$ and $Q_{{\rm sca}\,(2)}^{(0)}$ vanish, see the inset in Fig.~\ref{fig:Q0}. On the other hand, if $a_0$ and $a_2$ are replaced by $a_0^{(\rm PEC)}$ and $a_2^{(\rm PEC)}$, respectively, which should correspond to the beginning of the scattering, when the two resonant partitions are not excited yet, $Q_{{\rm sca}\,(0)}^{(0)}$ equals 1.290... Thus, all three aforementioned conditions are fulfilled.

\emph{Dimensionless variables.} Let $\omega_{_F}$ be the frequency, which for a given $R$ corresponds to $q_{_F}$. Then, \mbox{$\omega = \omega_{_F}q/q_{_F}$} and it is convenient to introduce the dimensionless time equal to $\omega_{_F} t$.

\emph{Computer Simulation.} For the detailed study of the dynamical scattering, we perform the direct numerical integration of the complete set of the Maxwell equations supplemented by the usual boundary conditions~\cite{bohren2008absorption}. The simulation is performed with the help of\emph{ Lumerical FDTD solver}.  The pulse duration $\tau = 329/\omega_{_F}$.

The simulation results are shown in Fig.~\ref{fig:Q0}. In complete agreement with the stipulated above generic features of the DFRs, $Q_{\rm sca}(t)$ exhibits two sharp intensive flashes --- at the beginning of the scattering process and after the incident pulse is over. The temporal dependence of the corresponding scattering diagram is shown in a movie presented in the Supplemental Material~II~\footnote{See the Supplemental Material~II at http://link.aps.org/supplemental/???/PhysRevLett.??? for the dynamics of the scattering diagram. The time is presented in units $\omega_{_F}t$.}. Note the qualitative changes in the directionality of the scattering radiation during the transients.

However, the toy model of a \emph{single} driven oscillator with a complex eigenfrequency provides the excellent quantitative description of the problem too. Within the framework of the discussed adiabatic approximation, we may suppose that the scattered filed pattern is still described by the formulae for the steady-state scattering but with the time-dependent scattering coefficients. For the off-resonant coefficients and background partitions (PEC modes) this dependence is reduced to the factor $\exp[-i\omega_{_F}t]$. The only difference for the resonant modes, i.e., for $d_0(t)$ and $d_{\pm 2}(t)$, is that their steady-state values are multiplied by $\exp[-(\gamma_\ell+i\omega_{_F})t]$, where $\gamma_\ell$ is the half-width of the corresponding steady-state resonant profile. The comparison of the $Q_{\rm sca}$ obtained in this approach with the results of the direct numerical integration of the Maxwell equations is shown in Fig.~\ref{fig:Q0}.

\emph{Conclusions.} Summarizing the obtained results we emphasize that the generic property of the DFRs is fast dynamics of the background partitions and slow dynamics of the resonant ones. The characteristic time scale of the slow dynamics equals the inverse linewidth of the resonant mode in the steady state. The larger the $Q$-factor of this mode, the slower the transient dynamics. This property gives rise to qualitatively different manifestations of the DRFs relative to the steady-state case. Specifically, if the exciting external pulse have a sharp variation of its amplitude with the characteristic time scale of the variation smaller then the transient period for the slow dynamics, the conditions for the complete destructive interference are not fulfilled at the beginning of the transient process caused by the variation. It gives rise to intensive ``flashes" of the output signal, no matter whether the amplitude of the exciting signal increases or decreases. In particular, the flash should be exhibited beyond the trailing edge of the exciting pulse \emph{after the exciting pulse is already over}. The proposed toy model, where the amplitudes of the off-resonant modes and background partitions are supposed to reach the steady-state values instantaneously, while the slow dynamics of the resonant partitions is described by the ones of the driven oscillators with complex eigenfrequencies provides the excellent quantitative description of the phenomenon. It should be stressed that in addition to the description of the integral characteristics (such as cross sections) the model describes the evolution of the the directional scattering and even the fine structure of the entire scattered field (including the birth and annihilation of its singular points).

We believe this study shed a new light to the famous and well-known problem and could find plenty of new applications in nanooptics (e.g., for generating ultrashort scattered pulses by a nanoparticle, a new way to manipulate the intensity and directionality of the scattered radiation, as elements of optical computers~\cite{silva2014performing}, etc.) and related fields.
%
\begin{acknowledgments}
M.I.T. acknowledges the financial support of Russian
Foundation for Basic Research (Grant No. 17-02-00401)
for the analytical study and  Russian Science Foundation
(Grant No. 14-19-01599) for the computer simulation. The modeling of the resonant light scattering was supported by the MEPhI Academic Excellence Project (agreement with the Ministry of Education and Science of the Russian Federation of August 27, 2013, project no. 02.a03.21.0005). The work of A.E.M. was supported by the Australian Research Council and UNSW Scientia Fellowship.
\end{acknowledgments}

\bibliography{DF}

\begin{thebibliography}{22}%
\makeatletter
\providecommand \@ifxundefined [1]{%
 \@ifx{#1\undefined}
}%
\providecommand \@ifnum [1]{%
 \ifnum #1\expandafter \@firstoftwo
 \else \expandafter \@secondoftwo
 \fi
}%
\providecommand \@ifx [1]{%
 \ifx #1\expandafter \@firstoftwo
 \else \expandafter \@secondoftwo
 \fi
}%
\providecommand \natexlab [1]{#1}%
\providecommand \enquote  [1]{``#1''}%
\providecommand \bibnamefont  [1]{#1}%
\providecommand \bibfnamefont [1]{#1}%
\providecommand \citenamefont [1]{#1}%
\providecommand \href@noop [0]{\@secondoftwo}%
\providecommand \href [0]{\begingroup \@sanitize@url \@href}%
\providecommand \@href[1]{\@@startlink{#1}\@@href}%
\providecommand \@@href[1]{\endgroup#1\@@endlink}%
\providecommand \@sanitize@url [0]{\catcode `\\12\catcode `\$12\catcode
  `\&12\catcode `\#12\catcode `\^12\catcode `\_12\catcode `\%12\relax}%
\providecommand \@@startlink[1]{}%
\providecommand \@@endlink[0]{}%
\providecommand \url  [0]{\begingroup\@sanitize@url \@url }%
\providecommand \@url [1]{\endgroup\@href {#1}{\urlprefix }}%
\providecommand \urlprefix  [0]{URL }%
\providecommand \Eprint [0]{\href }%
\providecommand \doibase [0]{http://dx.doi.org/}%
\providecommand \selectlanguage [0]{\@gobble}%
\providecommand \bibinfo  [0]{\@secondoftwo}%
\providecommand \bibfield  [0]{\@secondoftwo}%
\providecommand \translation [1]{[#1]}%
\providecommand \BibitemOpen [0]{}%
\providecommand \bibitemStop [0]{}%
\providecommand \bibitemNoStop [0]{.\EOS\space}%
\providecommand \EOS [0]{\spacefactor3000\relax}%
\providecommand \BibitemShut  [1]{\csname bibitem#1\endcsname}%
\let\auto@bib@innerbib\@empty
\bibitem [{\citenamefont {Fano}(1961)}]{Fano:PR:1961}%
  \BibitemOpen
  \bibfield  {author} {\bibinfo {author} {\bibfnamefont {U.}~\bibnamefont
  {Fano}},\ }\href@noop {} {\bibfield  {journal} {\bibinfo  {journal} {Phys.
  Rev.}\ }\textbf {\bibinfo {volume} {124}},\ \bibinfo {pages} {1866} (\bibinfo
  {year} {1961})}\BibitemShut {NoStop}%
\bibitem [{\citenamefont {Luk'yanchuk}\ \emph {et~al.}(2010)\citenamefont
  {Luk'yanchuk}, \citenamefont {Zheludev}, \citenamefont {Maier}, \citenamefont
  {Halas}, \citenamefont {Nordlander}, \citenamefont {Giessen},\ and\
  \citenamefont {Chong}}]{Lukyanchuk2010}%
  \BibitemOpen
  \bibfield  {author} {\bibinfo {author} {\bibfnamefont {B.}~\bibnamefont
  {Luk'yanchuk}}, \bibinfo {author} {\bibfnamefont {N.~I.}\ \bibnamefont
  {Zheludev}}, \bibinfo {author} {\bibfnamefont {S.~A.}\ \bibnamefont {Maier}},
  \bibinfo {author} {\bibfnamefont {N.~J.}\ \bibnamefont {Halas}}, \bibinfo
  {author} {\bibfnamefont {P.}~\bibnamefont {Nordlander}}, \bibinfo {author}
  {\bibfnamefont {H.}~\bibnamefont {Giessen}}, \ and\ \bibinfo {author}
  {\bibfnamefont {C.~T.}\ \bibnamefont {Chong}},\ }\href
  {http://dx.doi.org/10.1038/nmat2810} {\bibfield  {journal} {\bibinfo
  {journal} {Nature Materials}\ }\textbf {\bibinfo {volume} {9}},\ \bibinfo
  {pages} {707} (\bibinfo {year} {2010})}\BibitemShut {NoStop}%
\bibitem [{\citenamefont {Miroshnichenko}\ \emph {et~al.}(2010)\citenamefont
  {Miroshnichenko}, \citenamefont {Flach},\ and\ \citenamefont
  {Kivshar}}]{Miroshnichenko2010}%
  \BibitemOpen
  \bibfield  {author} {\bibinfo {author} {\bibfnamefont {A.~E.}\ \bibnamefont
  {Miroshnichenko}}, \bibinfo {author} {\bibfnamefont {S.}~\bibnamefont
  {Flach}}, \ and\ \bibinfo {author} {\bibfnamefont {Y.~S.}\ \bibnamefont
  {Kivshar}},\ }\href {\doibase 10.1103/RevModPhys.82.2257} {\bibfield
  {journal} {\bibinfo  {journal} {Rev. Mod. Phys.}\ }\textbf {\bibinfo {volume}
  {82}},\ \bibinfo {pages} {2257} (\bibinfo {year} {2010})}\BibitemShut
  {NoStop}%
\bibitem [{\citenamefont {Jahani}\ and\ \citenamefont
  {Jacob}(2016)}]{Jacob:NN:2016}%
  \BibitemOpen
  \bibfield  {author} {\bibinfo {author} {\bibfnamefont {S.}~\bibnamefont
  {Jahani}}\ and\ \bibinfo {author} {\bibfnamefont {Z.}~\bibnamefont {Jacob}},\
  }\href {\doibase 10.1038/nnano.2015.304} {\bibfield  {journal} {\bibinfo
  {journal} {Nature Nanotechnology}\ }\textbf {\bibinfo {volume} {11}},\
  \bibinfo {pages} {23} (\bibinfo {year} {2016})}\BibitemShut {NoStop}%
\bibitem [{\citenamefont {Ott}\ \emph {et~al.}(2013)\citenamefont {Ott},
  \citenamefont {Kaldun}, \citenamefont {Raith}, \citenamefont {Meyer},
  \citenamefont {Laux}, \citenamefont {Evers}, \citenamefont {Keitel},
  \citenamefont {Greene},\ and\ \citenamefont {Pfeifer}}]{ott2013lorentz}%
  \BibitemOpen
  \bibfield  {author} {\bibinfo {author} {\bibfnamefont {C.}~\bibnamefont
  {Ott}}, \bibinfo {author} {\bibfnamefont {A.}~\bibnamefont {Kaldun}},
  \bibinfo {author} {\bibfnamefont {P.}~\bibnamefont {Raith}}, \bibinfo
  {author} {\bibfnamefont {K.}~\bibnamefont {Meyer}}, \bibinfo {author}
  {\bibfnamefont {M.}~\bibnamefont {Laux}}, \bibinfo {author} {\bibfnamefont
  {J.}~\bibnamefont {Evers}}, \bibinfo {author} {\bibfnamefont {C.~H.}\
  \bibnamefont {Keitel}}, \bibinfo {author} {\bibfnamefont {C.~H.}\
  \bibnamefont {Greene}}, \ and\ \bibinfo {author} {\bibfnamefont
  {T.}~\bibnamefont {Pfeifer}},\ }\href {\doibase 10.1126/science.1234407}
  {\bibfield  {journal} {\bibinfo  {journal} {Science}\ }\textbf {\bibinfo
  {volume} {340}},\ \bibinfo {pages} {716} (\bibinfo {year}
  {2013})}\BibitemShut {NoStop}%
\bibitem [{\citenamefont {Kaldun}\ \emph {et~al.}(2016)\citenamefont {Kaldun},
  \citenamefont {Bl{\"a}ttermann}, \citenamefont {Stoo{\ss}}, \citenamefont
  {Donsa}, \citenamefont {Wei}, \citenamefont {Pazourek}, \citenamefont
  {Nagele}, \citenamefont {Ott}, \citenamefont {Lin}, \citenamefont
  {Burgd{\"o}rfer} \emph {et~al.}}]{kaldun2016observing}%
  \BibitemOpen
  \bibfield  {author} {\bibinfo {author} {\bibfnamefont {A.}~\bibnamefont
  {Kaldun}}, \bibinfo {author} {\bibfnamefont {A.}~\bibnamefont
  {Bl{\"a}ttermann}}, \bibinfo {author} {\bibfnamefont {V.}~\bibnamefont
  {Stoo{\ss}}}, \bibinfo {author} {\bibfnamefont {S.}~\bibnamefont {Donsa}},
  \bibinfo {author} {\bibfnamefont {H.}~\bibnamefont {Wei}}, \bibinfo {author}
  {\bibfnamefont {R.}~\bibnamefont {Pazourek}}, \bibinfo {author}
  {\bibfnamefont {S.}~\bibnamefont {Nagele}}, \bibinfo {author} {\bibfnamefont
  {C.}~\bibnamefont {Ott}}, \bibinfo {author} {\bibfnamefont {C.}~\bibnamefont
  {Lin}}, \bibinfo {author} {\bibfnamefont {J.}~\bibnamefont {Burgd{\"o}rfer}},
   \emph {et~al.},\ }\href {\doibase 10.1126/science.aah6972} {\bibfield
  {journal} {\bibinfo  {journal} {Science}\ }\textbf {\bibinfo {volume}
  {354}},\ \bibinfo {pages} {738} (\bibinfo {year} {2016})}\BibitemShut
  {NoStop}%
\bibitem [{\citenamefont {Gruson}\ \emph {et~al.}(2016)\citenamefont {Gruson},
  \citenamefont {Barreau}, \citenamefont {Jim{\'e}nez-Galan}, \citenamefont
  {Risoud}, \citenamefont {Caillat}, \citenamefont {Maquet}, \citenamefont
  {Carr{\'e}}, \citenamefont {Lepetit}, \citenamefont {Hergott}, \citenamefont
  {Ruchon} \emph {et~al.}}]{gruson2016attosecond}%
  \BibitemOpen
  \bibfield  {author} {\bibinfo {author} {\bibfnamefont {V.}~\bibnamefont
  {Gruson}}, \bibinfo {author} {\bibfnamefont {L.}~\bibnamefont {Barreau}},
  \bibinfo {author} {\bibfnamefont {{\'A}.}~\bibnamefont {Jim{\'e}nez-Galan}},
  \bibinfo {author} {\bibfnamefont {F.}~\bibnamefont {Risoud}}, \bibinfo
  {author} {\bibfnamefont {J.}~\bibnamefont {Caillat}}, \bibinfo {author}
  {\bibfnamefont {A.}~\bibnamefont {Maquet}}, \bibinfo {author} {\bibfnamefont
  {B.}~\bibnamefont {Carr{\'e}}}, \bibinfo {author} {\bibfnamefont
  {F.}~\bibnamefont {Lepetit}}, \bibinfo {author} {\bibfnamefont {J.-F.}\
  \bibnamefont {Hergott}}, \bibinfo {author} {\bibfnamefont {T.}~\bibnamefont
  {Ruchon}},  \emph {et~al.},\ }\href {\doibase 10.1126/science.aah5188}
  {\bibfield  {journal} {\bibinfo  {journal} {Science}\ }\textbf {\bibinfo
  {volume} {354}},\ \bibinfo {pages} {734} (\bibinfo {year}
  {2016})}\BibitemShut {NoStop}%
\bibitem [{\citenamefont {Sirenko}\ \emph {et~al.}(2007)\citenamefont
  {Sirenko}, \citenamefont {Str{\"o}m},\ and\ \citenamefont
  {Yashina}}]{sirenko2007modeling}%
  \BibitemOpen
  \bibfield  {author} {\bibinfo {author} {\bibfnamefont {Y.~K.}\ \bibnamefont
  {Sirenko}}, \bibinfo {author} {\bibfnamefont {S.}~\bibnamefont {Str{\"o}m}},
  \ and\ \bibinfo {author} {\bibfnamefont {N.~P.}\ \bibnamefont {Yashina}},\
  }\href@noop {} {\emph {\bibinfo {title} {Modeling and analysis of transient
  processes in open resonant structures: New methods and techniques}}},\ Vol.\
  \bibinfo {volume} {122}\ (\bibinfo  {publisher} {Springer},\ \bibinfo {year}
  {2007})\BibitemShut {NoStop}%
\bibitem [{\citenamefont {{Golovinski}}\ \emph {et~al.}(2017)\citenamefont
  {{Golovinski}}, \citenamefont {{Yakovets}},\ and\ \citenamefont
  {{Khramov}}}]{Golovinski:ArXiv2017}%
  \BibitemOpen
  \bibfield  {author} {\bibinfo {author} {\bibfnamefont {P.~A.}\ \bibnamefont
  {{Golovinski}}}, \bibinfo {author} {\bibfnamefont {A.~V.}\ \bibnamefont
  {{Yakovets}}}, \ and\ \bibinfo {author} {\bibfnamefont {E.~S.}\ \bibnamefont
  {{Khramov}}},\ }\href@noop {} {\bibfield  {journal} {\bibinfo  {journal}
  {ArXiv e-prints}\ } (\bibinfo {year} {2017})},\ \Eprint
  {http://arxiv.org/abs/1711.02498} {arXiv:1711.02498 [physics.optics]}
  \BibitemShut {NoStop}%
\bibitem [{\citenamefont {Polyanskiy}()}]{refractiveindex.info}%
  \BibitemOpen
  \bibfield  {author} {\bibinfo {author} {\bibfnamefont {M.}~\bibnamefont
  {Polyanskiy}},\ }\href {http://refractiveindex.info/} {\emph {\bibinfo
  {title} {Refractive index database}}},\ \bibinfo {note}
  {http://refractiveindex.info/}\BibitemShut {NoStop}%
\bibitem [{\citenamefont {Joe}\ \emph {et~al.}(2006)\citenamefont {Joe},
  \citenamefont {Satanin},\ and\ \citenamefont {Kim}}]{Joe2006}%
  \BibitemOpen
  \bibfield  {author} {\bibinfo {author} {\bibfnamefont {Y.~S.}\ \bibnamefont
  {Joe}}, \bibinfo {author} {\bibfnamefont {A.~M.}\ \bibnamefont {Satanin}}, \
  and\ \bibinfo {author} {\bibfnamefont {C.~S.}\ \bibnamefont {Kim}},\ }\href
  {\doibase 10.1088/0031-8949/74/2/020} {\bibfield  {journal} {\bibinfo
  {journal} {Phys. Scr.}\ }\textbf {\bibinfo {volume} {74}},\ \bibinfo {pages}
  {259} (\bibinfo {year} {2006})},\ \Eprint {http://arxiv.org/abs/0111100v1}
  {arXiv:0111100v1 [arXiv:physics]} \BibitemShut {NoStop}%
\bibitem [{\citenamefont {Tribelsky}(2014)}]{Tribelsky:RITS}%
  \BibitemOpen
  \bibfield  {author} {\bibinfo {author} {\bibfnamefont {M.~I.}\ \bibnamefont
  {Tribelsky}},\ }\href
  {https://www.researchgate.net/publication/273693137_Linear_and_Nonlinear_Evolution_in_Time_and_Space}
  {\emph {\bibinfo {title} {Linear and Nonlinear Evolution in Time and
  Space}}}\ (\bibinfo  {publisher} {Shigemasa Printing, Yamaguchi},\ \bibinfo
  {year} {2014})\ \bibinfo {note}
  {https://www.researchgate.net/publication/273693137\\\_Linear\_and\_Nonlinear\_Evolution\_in\_Time\_and\_Space}\BibitemShut
  {NoStop}%
\bibitem [{\citenamefont {Louisell}(1960)}]{louisell1960coupled}%
  \BibitemOpen
  \bibfield  {author} {\bibinfo {author} {\bibfnamefont {W.~H.}\ \bibnamefont
  {Louisell}},\ }\href@noop {} {\emph {\bibinfo {title} {Coupled mode and
  parametric electronics}}}\ (\bibinfo  {publisher} {Wiley},\ \bibinfo {year}
  {1960})\BibitemShut {NoStop}%
\bibitem [{\citenamefont {Ruan}\ and\ \citenamefont
  {Fan}(2010{\natexlab{a}})}]{Fan:PRL:2010}%
  \BibitemOpen
  \bibfield  {author} {\bibinfo {author} {\bibfnamefont {Z.}~\bibnamefont
  {Ruan}}\ and\ \bibinfo {author} {\bibfnamefont {S.}~\bibnamefont {Fan}},\
  }\href {\doibase 10.1103/PhysRevLett.105.013901} {\bibfield  {journal}
  {\bibinfo  {journal} {Phys. Rev. Lett.}\ }\textbf {\bibinfo {volume} {105}},\
  \bibinfo {pages} {013901} (\bibinfo {year} {2010}{\natexlab{a}})}\BibitemShut
  {NoStop}%
\bibitem [{\citenamefont {Verslegers}\ \emph {et~al.}(2010)\citenamefont
  {Verslegers}, \citenamefont {Yu}, \citenamefont {Catrysse},\ and\
  \citenamefont {Fan}}]{Verslegers:10}%
  \BibitemOpen
  \bibfield  {author} {\bibinfo {author} {\bibfnamefont {L.}~\bibnamefont
  {Verslegers}}, \bibinfo {author} {\bibfnamefont {Z.}~\bibnamefont {Yu}},
  \bibinfo {author} {\bibfnamefont {P.~B.}\ \bibnamefont {Catrysse}}, \ and\
  \bibinfo {author} {\bibfnamefont {S.}~\bibnamefont {Fan}},\ }\href {\doibase
  10.1364/JOSAB.27.001947} {\bibfield  {journal} {\bibinfo  {journal} {J. Opt.
  Soc. Am. B}\ }\textbf {\bibinfo {volume} {27}},\ \bibinfo {pages} {1947}
  (\bibinfo {year} {2010})}\BibitemShut {NoStop}%
\bibitem [{\citenamefont {Ruan}\ and\ \citenamefont
  {Fan}(2010{\natexlab{b}})}]{Fan:jp9089722}%
  \BibitemOpen
  \bibfield  {author} {\bibinfo {author} {\bibfnamefont {Z.}~\bibnamefont
  {Ruan}}\ and\ \bibinfo {author} {\bibfnamefont {S.}~\bibnamefont {Fan}},\
  }\href {\doibase 10.1021/jp9089722} {\bibfield  {journal} {\bibinfo
  {journal} {The Journal of Physical Chemistry C}\ }\textbf {\bibinfo {volume}
  {114}},\ \bibinfo {pages} {7324} (\bibinfo {year}
  {2010}{\natexlab{b}})}\BibitemShut {NoStop}%
\bibitem [{\citenamefont {Hsu}\ \emph {et~al.}(2014)\citenamefont {Hsu},
  \citenamefont {DeLacy}, \citenamefont {Johnson}, \citenamefont
  {Joannopoulos},\ and\ \citenamefont {Soljacic}}]{Soljacic:nl500340n}%
  \BibitemOpen
  \bibfield  {author} {\bibinfo {author} {\bibfnamefont {C.~W.}\ \bibnamefont
  {Hsu}}, \bibinfo {author} {\bibfnamefont {B.~G.}\ \bibnamefont {DeLacy}},
  \bibinfo {author} {\bibfnamefont {S.~G.}\ \bibnamefont {Johnson}}, \bibinfo
  {author} {\bibfnamefont {J.~D.}\ \bibnamefont {Joannopoulos}}, \ and\
  \bibinfo {author} {\bibfnamefont {M.}~\bibnamefont {Soljacic}},\ }\href
  {\doibase 10.1021/nl500340n} {\bibfield  {journal} {\bibinfo  {journal} {Nano
  Letters}\ }\textbf {\bibinfo {volume} {14}},\ \bibinfo {pages} {2783}
  (\bibinfo {year} {2014})}\BibitemShut {NoStop}%
\bibitem [{Note1()}]{Note1}%
  \BibitemOpen
  \bibinfo {note} {See the Supplemental Material I at
  http://link.aps.org/supplemental/???/PhysRevLett.??? for details of routine
  cumbersome calculations.}\BibitemShut {Stop}%
\bibitem [{\citenamefont {Bohren}\ and\ \citenamefont
  {Huffman}(2008)}]{bohren2008absorption}%
  \BibitemOpen
  \bibfield  {author} {\bibinfo {author} {\bibfnamefont {C.~F.}\ \bibnamefont
  {Bohren}}\ and\ \bibinfo {author} {\bibfnamefont {D.~R.}\ \bibnamefont
  {Huffman}},\ }\href@noop {} {\emph {\bibinfo {title} {Absorption and
  scattering of light by small particles}}}\ (\bibinfo  {publisher} {John Wiley
  \& Sons},\ \bibinfo {year} {2008})\BibitemShut {NoStop}%
\bibitem [{\citenamefont {Tribelsky}\ and\ \citenamefont
  {Miroshnichenko}(2016)}]{Tribelsky_Mirosh_2016}%
  \BibitemOpen
  \bibfield  {author} {\bibinfo {author} {\bibfnamefont {M.~I.}\ \bibnamefont
  {Tribelsky}}\ and\ \bibinfo {author} {\bibfnamefont {A.~E.}\ \bibnamefont
  {Miroshnichenko}},\ }\href {\doibase 10.1103/physreva.93.053837} {\bibfield
  {journal} {\bibinfo  {journal} {Physical Review A}\ }\textbf {\bibinfo
  {volume} {93}},\ \bibinfo {pages} {053837} (\bibinfo {year}
  {2016})}\BibitemShut {NoStop}%
\bibitem [{Note2()}]{Note2}%
  \BibitemOpen
  \bibinfo {note} {See the Supplemental Material~II at
  http://link.aps.org/supplemental/???/PhysRevLett.??? for the dynamics of the
  scattering diagram. The time is presented in units $\omega
  _{_F}t$.}\BibitemShut {Stop}%
\bibitem [{\citenamefont {Silva}\ \emph {et~al.}(2014)\citenamefont {Silva},
  \citenamefont {Monticone}, \citenamefont {Castaldi}, \citenamefont {Galdi},
  \citenamefont {Al{\`u}},\ and\ \citenamefont
  {Engheta}}]{silva2014performing}%
  \BibitemOpen
  \bibfield  {author} {\bibinfo {author} {\bibfnamefont {A.}~\bibnamefont
  {Silva}}, \bibinfo {author} {\bibfnamefont {F.}~\bibnamefont {Monticone}},
  \bibinfo {author} {\bibfnamefont {G.}~\bibnamefont {Castaldi}}, \bibinfo
  {author} {\bibfnamefont {V.}~\bibnamefont {Galdi}}, \bibinfo {author}
  {\bibfnamefont {A.}~\bibnamefont {Al{\`u}}}, \ and\ \bibinfo {author}
  {\bibfnamefont {N.}~\bibnamefont {Engheta}},\ }\href {\doibase
  10.1126/science.1242818} {\bibfield  {journal} {\bibinfo  {journal}
  {Science}\ }\textbf {\bibinfo {volume} {343}},\ \bibinfo {pages} {160}
  (\bibinfo {year} {2014})}\BibitemShut {NoStop}%
\end{thebibliography}%
\end{document}